\documentclass{article}
\usepackage{amsmath}
\usepackage{geometry}
\usepackage{graphicx}
\newgeometry{vmargin={15mm}, hmargin={12mm,17mm}}

\title{A Unique Approach to Classify Inflationary Potentials}
\author{Somnath Das\footnote{20phph12@uohyd.ac.in} and P K Suresh\footnote{sureshpk@uohyd.ac.in}\\
\emph{School of Physics, University of Hyderabad}\\
\emph{ Central University P.O, Hyderabad - 500046. India.}}

\date{}

\begin{document}

\maketitle
\begin{abstract}
    Inflationary cosmology has made significant strides in understanding the physics driving the rapid expansion of the early universe. However, many inflation models with diverse potential shapes present analysis, comparison, and classification challenges. In this paper, we propose a novel approach to tackle this issue. We introduce a general potential formula encompassing all inflationary potentials, whether single-field or multi-field, into a single mathematical framework. This formula establishes a unified framework for systematically classifying inflation models based on their potential functions. We showcase the efficacy of the general potential formula by successfully reproducing well-known inflation models, such as the Starobinsky potential and the Valley Hybrid Inflation model. Moreover, we derive general inflationary parameters, including the slow-roll parameters and power spectra, using the proposed formula. Our approach provides a versatile tool for classifying and studying various inflationary scenarios, simplifying the analysis and comparison of different models in the field of inflationary cosmology.
\end{abstract}

\section{\label{sec:level1}Introduction}
The study of inflationary cosmology has revolutionized our understanding of the early universe, providing a compelling framework to explain the observed isotropy, homogeneity, and flatness \cite{PhysRevD.23.347}. Inflationary models posit a period of rapid expansion in the early universe, driven by a homogenous scalar field known as inflaton    \cite{Baumann:2014nda}. These models have successfully addressed the shortcomings of the standard Big Bang cosmology, such as the horizon and flatness problems, and have provided predictions that align with a wide range of cosmological observations \cite{Baumann:2014nda}. While inflationary cosmology has made remarkable progress, the field remains rich with many inflation models, each characterized by different potentials for inflation. The diversity of these potentials arises from the underlying physics and the dynamics of the early universe, giving rise to many inflationary scenarios. This variety poses challenges in analyzing, comparing, and classifying different models, hindering our ability to gain deeper insights into the fundamental physics driving inflation. Inflationary cosmology has witnessed numerous efforts to classify and categorize inflation models based on their underlying physics and potential shapes. These classification schemes aim to capture the diversity of inflationary potentials and provide a systematic framework for understanding the range of inflation scenarios proposed in the literature. 

One common approach to classifying inflationary models is based on the shape of the inflation potential. This method categorizes inflation potentials into power-law potentials, exponential potentials, hybrid potentials, \cite{Baumann:2014nda} etc. Each class represents a specific functional form for the potential, often motivated by specific underlying physics or symmetry considerations. While this approach provides a straightforward categorization, it can be limited in its ability to capture the full diversity of inflation models, as it may overlook subtle variations within each class. 
Another classification method involves characterizing inflationary models based on their observational predictions. This approach considers inflation predictions for various cosmological observables, such as the spectral index of primordial fluctuations, the tensor-to-scalar ratio, and the non-Gaussianity parameter \cite{Baumann:2014nda}. Models that yield similar predictions for these observables are grouped, forming a classification based on the resulting observational signatures \cite{PotCom}. This method is valuable for connecting inflation models to empirical data such as the CMB observations. Still, it may need to fully capture the potential shapes that give rise to these predictions. 

In addition to the classification schemes based on the shape of the inflation potential and observational predictions, another important categorization in the field of inflationary cosmology is based on the energy scales involved in the inflationary process. This classification distinguishes high-field and low-field inflation models \cite{PotCom}. Along this line, various physical properties might be considered to classify the inflationary potentials. In this paper, we propose a novel approach to address this challenge by presenting a general potential formula that encompasses all inflationary potentials. By allowing the functional form of the potential to be chosen, our approach offers flexibility in capturing the diversity of inflationary scenarios. It unifies all the inflationary potentials, single-field or multi-field, under one functional form.

\section{General Potential}\label{sec:GP}
The following comprehensive general formula encapsulates the dynamics of inflationary models with multiple fields. Our formula offers a unified approach, enabling the seamless incorporation of any inflationary model into a single mathematical framework. 

\begin{equation}
    V(\varphi_{m}) = M^4 \bar{A}(\varphi_{m}) \left[\bar{B}(\varphi_{m}) + \sum_{i=0}^{n}a_{i}\bar{C_{i}}(\varphi_{m})\right]^p. \label{genroot}
\end{equation}

Here, $a_{0} = 0$. $M$ represents a characteristic energy scale associated with the inflationary dynamics in the inflation potential. It sets the overall magnitude of the potential energy during inflation. Equation (\ref{genroot}) offers a unified framework for systematically classifying inflation models. The potential, denoted as $V(\varphi_{m})$, is expressed in terms of a set of functions $\bar{A}(\varphi_{m})$, $\bar{B}(\varphi_{m})$, and $\bar{C_i}(\varphi_{m})$, where $i$ ranges from $0$ to $n$ and $p$ is any real number. Here, $m= 0,1,2..t$ where $t$ is any integer. $m=0$ corresponds to the inflaton field $\phi$; other $m$ values can represent other fields. If $m$ is single-valued, it represents single-field inflation, and multi-valued $m$ represents multifield models. We may classify inflationary models based on the value of $n$, i.e., up to the number $i$ runs in the summation. We can imagine a series of potentials with various possible functions for each value. In equation (\ref{genroot}), particular $n$ values correspond to the number of terms linearly associated with the potential, $n=0$ corresponds to single-termed potentials, $n=1$ corresponds to double-termed potentials, and so on. For the single field inflation model, $m = 0$, and the potential is a function of $\varphi_{o} = \phi$ alone.

\begin{equation}
    V(\phi) = M^4A(\phi)\left[B(\phi) + \sum_{i=0}^{n}a_{i}C_{i}(\phi)\right]^{p}. \label{root}
\end{equation}

To demonstrate the effectiveness of the general potential formula for a single field (\ref{root}), we will show how a specific choice of the functions and parameters within the formula reproduces the well-known Starobinsky inflation model. By selecting $n=1$, $A(\phi) = B(\phi) = 1$, $a_{1} = -1$, $C_{1}(\phi) = e^{-\sqrt{\frac{2}{3}}\frac{\phi}{M_{\text{Pl}}}}$, and $p=2$, we can obtain the Starobinsky potential \cite{PotCom},\cite{STAROBINSKY198099}.

\begin{equation}
V(\phi) = M^4\left[1 - e^{-\sqrt{\frac{2}{3}}\frac{\phi}{M_{\text{Pl}}}}\right]^2. \label{star}
\end{equation}

Further, in the equation (\ref{genroot}) if we choose $n=4$ with $\bar{A}(\varphi_m) = 1, \bar{B}(\varphi_m) = \frac{\lambda_{\sigma}}{4}, \bar{C_{1}}(\varphi_m) = \sigma^4, \bar{C_{2}}(\varphi_m) = \sigma^2, \bar{C_{3}}(\varphi_m) = \phi^2, \bar{C_{4}}(\varphi_m) = \phi^2 \sigma^2$ with coefficients $a_{1} = \frac{\lambda_{\sigma}}{4M^4}, a_{2} = -\frac{\lambda_{\sigma}}{2M^2}, a_{3} = \frac{m_{\phi}^2}{2M^4}, a_{4} = \frac{\lambda}{2M^4}$, $m=0,1$ with $\varphi_{0} = \phi$ and $\varphi_1 = \sigma$, it reproduces the Valley Hybrid Inflation model \cite{PhysRevD.49.748}, \cite{Malsawmtluangi_2021}.

\begin{equation}
    V(\phi, \sigma) = \frac{\lambda_{\sigma}}{4}(\sigma^2 - M^2)^2 + \frac{1}{2}m_{\phi}^2\phi^2 + \frac{1}{2}\lambda\phi^2 \sigma^2. \label{vhi}
\end{equation}

Also, in equation (\ref{genroot}), if we consider $n=1$, $A(\phi) = -\frac{3}{8l^2}\phi^2$, $B(\phi) = 1$, $a_{1} = -\frac{4\kappa l^2}{9}$, $C_{1} = \phi^4$, and $p=1$, it is easy to see that one attains the effective potential that comes from the AdS Swampland conjectures \cite{ALVAREZGARCIA2022136861}.

\begin{equation}
    V(\phi) = -\frac{3}{8l^2}\phi^2 + \frac{\kappa}{6}\phi^6.
\end{equation}

It's apparent that the general potential formulation is highly versatile and accommodates a wide range of inflationary potentials. It provides a flexible framework to describe various inflationary models and allows for exploring different inflationary dynamics.

\section{Inflationary Parameters}
This section explores the implications of the proposed general potential formula for deriving general inflationary parameters that characterize inflationary models, as outlined in equation (\ref{root}). By analyzing the functional form of the potential and its corresponding dynamics, we can extract essential parameters that quantify the behavior and observational predictions of inflation. We drop the scalar field $\phi$ from the expression to ensure better readability. 

The general form of the first slow-roll parameter is

\begin{align}
\epsilon &= \frac{p^2}{16\pi}\bigg[A^{\frac{1}{p}}B + \sum_{i=0}^{n}a_{i}C_{i}A^{\frac{1}{p}}\bigg]^{-2} \bigg[\frac{1}{p}A^{\frac{1}{p}-1}A^{'}B + A^{\frac{1}{p}}B^{'} + \sum_{i=0}^{n}a_{i}C_{i}^{'}A^{\frac{1}{p}} + \frac{1}{p}\sum_{i=0}^{n}a_{i}C_{i}A^{\frac{1}{p}-1}A^{'}\bigg]^2. \label{genep}
\end{align}

Here, $X^{'} = \frac{dX}{d\phi}$. The tensor-to-scalar ratio can be expressed using the first slow roll parameter using the relation, $r =16\epsilon$. This provides a dynamic approach for choosing the functions by reflecting the latest bound on the tensor-to-scalar ratio, a crucial observational quantity in inflationary cosmology. Recent measurements and cosmic microwave background radiation data analysis have constrained the tensor-to-scalar ratio. The current bound is $r<0.035$ \cite{Update}, which might get as low as $r<0.004$ with future observations \cite{10.1093/ptep/ptac150} \cite{Barron_2018}. Incorporating these bounds on $r$ is a valuable guideline while constructing inflationary models using the proposed general potential formula, allowing us to explore the range of viable scenarios within the context of current observational constraints. The general form of the second slow-roll parameter is,
\begin{equation}
\begin{aligned}
\eta = \frac{p}{8\pi} \bigg[&A^{\frac{1}{p}}B + \sum_{i=0}^{n}a_{i}C_{i}A^{\frac{1}{p}} \bigg]^{-2} \times \bigg[ (p-1)\Big(\frac{1}{p}A^{\frac{1}{p}-1}A^{'}B + A^{\frac{1}{p}}B^{'} + \sum_{i=0}^{n}a_{i}C_{i}^{'}A^{\frac{1}{p}} + \frac{1}{p}\sum_{i=0}^{n}a_{i}C_{i}A^{\frac{1}{p}-1}A^{'}\Big)^2 \\
&+ \bigg(A^{\frac{1}{p}}B + \sum_{i=0}^{n}a_{i}C_{i}A^{\frac{1}{p}}\bigg) \bigg[ \frac{1}{p}\big(\frac{1}{p}-1\big)A^{\frac{1}{p}-2}(A^{'})^2B + \frac{1}{p}A^{\frac{1}{p}-1}A^{''}B + \frac{2}{p}A^{\frac{1}{p}-1}A^{'}B^{'} + A^{\frac{1}{p}}B^{''} \\
&+ \sum_{i=0}^{n}a_{i}C_{i}^{''}A^{\frac{1}{p}} +\frac{2}{p}\sum_{i=0}^{n}a_{i}C_{i}^{'}A^{\frac{1}{p}-1}A^{'} +\frac{1}{p}\big(\frac{1}{p}-1\big)\sum_{i=0}^{n}a_{i}C_{i}A^{\frac{1}{p}-2}(A^{'})^2 + \frac{1}{p}\sum_{i=0}^{n}a_{i}C_{i}A^{\frac{1}{p}-1}A^{''}\bigg]\bigg]. \label{geneta}
\end{aligned}
\end{equation}

Let's consider the Starobinsky potential (\ref{star}) for testing the general slow roll parameters. Considering the particular functions mentioned in section \ref{sec:GP} applicable for generating Starobinsky potential, we use equation (\ref{genep}) to produce the first slow roll parameter $\epsilon_{s}$.

\begin{equation}
    \epsilon_{s} = \frac{1}{6\pi M_{pl}^2}e^{-\sqrt{\frac{2}{3}}\frac{2\phi}{M_{\text{pl}}}}\bigg[1 - e^{-\sqrt{\frac{2}{3}}\frac{\phi}{M_{\text{pl}}}}\bigg]^{-2}. \label{starep}
\end{equation}

Using the general formula for the second slow roll parameter (\ref{geneta}), one can find the second slow roll parameter for the Starobinsky potential $\eta_{s}$.

\begin{equation}
    \eta_{s} = \frac{1}{6\pi M_{pl}^{2}}e^{-\sqrt{\frac{2}{3}}\frac{\phi}{M_{\text{Pl}}}}\bigg[4e^{-\sqrt{\frac{2}{3}}\frac{\phi}{M_{\text{Pl}}}} -1\bigg].
\end{equation}

In this way, the general slow roll parameters can be used to find the slow roll parameters of any model. The potential (\ref{root}) can also generalize other inflationary parameters, such as the scalar power spectrum.

\begin{align}
    P_{S} =& \frac{128M^{4}}{3p^2} \bigg[A^{\frac{1}{p}}B + \sum_{i=0}^{n}a_{i}C_{i}A^{\frac{1}{p}}\bigg]^{p-2} \bigg[\frac{1}{p}A^{\frac{1}{p}}A^{'}B + A^{\frac{1}{p}}B^{'} +\sum_{i=0}^{n}a_{i}C_{i}^{'}A^{\frac{1}{p}} + \frac{1}{p}\sum_{i=0}^{n}a_{i}C_{i}A^{\frac{1}{p}-1}A^{'}\bigg]^2.
\end{align}

The tensor power spectrum can be generally expressed as
\begin{align}
    P_{T} =& \frac{128 A}{3}\bigg[B + \sum_{i=0}^{n}a_{i}C_{i}\bigg]^p.
\end{align}

The ratio of $P_{T}$ and $P_{S}$ gives us the tensor-to-scalar ratio $r$. We have derived the general forms of the inflationary parameters considering $m=0$, which essentially considers the single-field inflationary models. In the inflationary parameters, if we consider the functions of inflaton field $A(\phi), B(\phi)$ and $C_{i}(\phi)$ as multi-field functions such as $\bar{A}(\varphi_{m}), \bar{B}(\varphi_{m})$ and $\bar{C_{i}}(\varphi_{m})$ then these parameters will be valid for multi-field potentials as well. In such a scenario, the derivatives of the functions will be partial depending on the model in context.

Various other inflationary parameters can be obtained from the general potential, such as the other slow-roll parameters, the scalar spectral index, the running of the indices, and so on.

\section{Conclusion}
The general potential and associated parameters offer a flexible and comprehensive framework for analyzing various inflationary models. By employing the general potential, it becomes possible to classify existing inflationary models within a unified framework. Moreover, this framework provides a systematic approach to constructing new inflationary models based on functional considerations. One of the strengths of this general potential lies in its ability to generate an infinite number of inflationary potentials. The general form allows for a broad range of functional forms for $A(\phi)$, $B(\phi)$, and $C_i(\phi)$, enabling a vast exploration of potential shapes and behaviors. This flexibility is crucial for accommodating diverse inflationary dynamics from different underlying physics. It should be noted that the general potential is focused on inflation with scalar fields.

However, given the infinite possibilities, it becomes necessary to constrain the parameter space by selecting a subset of inflationary potentials that align with current observational data. The aim is to identify those potentials that successfully reproduce the observed cosmic microwave background radiation, primordial density fluctuations, and other relevant cosmological observables. Selecting a subset of inflationary potentials consistent with data narrows the range of viable models.
Through this systematic approach, the general potential provides a powerful tool for refining existing inflationary models and discovering new, relevant inflationary scenarios. Focusing on a functional basis allows us to explore unexplored regions of the parameter space and discover novel inflationary models that may exhibit unique features or address current theoretical or observational challenges.

\section*{Acknowledgement} S. D. acknowledges the financial support the Govt of India  through the Prime Minister's Research Fellowship (PMRF).


\begin{thebibliography}{99}
\bibitem{PhysRevD.23.347}
  A. H. Guth,
  \emph{Inflationary universe: A possible solution to the horizon and flatness problems},
  \emph{Phys. Rev. D} \textbf{23}, 347 (1981).

\bibitem{Baumann:2014nda}
  D. Baumann and L. McAllister,
  \emph{Inflation and String Theory},
  {arXiv:1404.2601 [hep-th]},
  CUP (2015).


\bibitem{PotCom}
  J.  Martin,
  \emph{The best inflationary models after Planck},
  \emph{JCAP} \textbf{03}, 039 (2014).

\bibitem{STAROBINSKY198099}
  A. A. Starobinsky,
  \emph{A new type of isotropic cosmological models without singularity},
  \emph{Phys. Lett. B}  \textbf{91}, 99 (1980).

\bibitem{PhysRevD.49.748}
  A. Linde,
  \emph{Hybrid inflation},
  \emph{Phys. Rev. D} \textbf{49}, 748 (1994).

\bibitem{Malsawmtluangi_2021}
  N. Malsawmtluangi,
  \emph{Analytical study of classic models of hybrid inflation},
  \emph{J. Phys. Commun.} \textbf{5}, 085016 (2021).

\bibitem{ALVAREZGARCIA2022136861}
  R. Álvarez-García \textit{et al.},
  \emph{Swampland conjectures for an almost topological gravity theory},
  \emph{Phys. Lett. B} \textbf{825}, 136861 (2022).

\bibitem{Update}
  M. Forconi \textit{et al.},
  \emph{Cosmological constraints on slow roll inflation: An update},
  \emph{Phys. Rev. D} \textbf{104}, 103528 (2022).

\bibitem{10.1093/ptep/ptac150}
  Collaboration LiteBIRD,
  \emph{Probing cosmic inflation with the LiteBIRD cosmic microwave background polarization survey},
  \emph{Prog. Theor. Exp. Phys.} \textbf{2023}, 4 (2022).

\bibitem{Barron_2018}
  D. Barron \textit{et al.},
  \emph{Optimization study for the experimental configuration of CMB-S4},
  \emph{JCAP} \textbf{02}, 009 (2018).




\end{thebibliography}
\end{document}